\documentclass[10pt]{article}
\usepackage[utf8]{inputenc}
\usepackage{amssymb}
\usepackage{amsmath}
\usepackage{graphicx}
\usepackage{xcolor}

\newcommand{\D}{\,\mathrm{d}}
\newcommand{\const}{\operatorname{const}}
\newcommand{\q}{\mathbf{q}}
\newcommand{\p}{\mathbf{p}}

\newcommand{\ket}[1]{\ensuremath{\left|#1\right>}}
\newcommand{\bra}[1]{\ensuremath{\left<#1\right|}}
\newcommand{\avrg}[1]{\ensuremath{\left<#1\right>}}
\DeclareMathOperator{\Tr}{Tr}

\newcommand{\E}{\mathrm{e}}
\newcommand{\I}{\mathrm{i}}

\begin{document}
\title{Yet Another Approach to Loschmidt's Paradox}
\author{Lev A.\,Melnikovsky\thanks{E-mail: leva@kapitza.ras.ru}\\
\\
\textit{\small P.L.\,Kapitza Institute for Physical Problems, Russian Academy of Sciences}}
\date{} 
\maketitle

\begin{abstract}
The works by Lev Petrovich Pitaevskii are reference points for choosing an interesting research topic. An example is the article \cite{pit} which promotes rigorous results in non-equilibrium statistical physics. In present paper, we rigorously prove that a non-equilibrium state, on the average, is a local entropy minimum. This statement corresponds to the ``entropy growth'' of statistical mechanics and does not violate time reversal symmetry of microscopic motion: the first-order time derivative of the entropy is zero $\dot{S}=0$, while the second order derivative is non-negative $\ddot{S}\ge 0$.
\end{abstract}

\section{Introduction}
According to the Second Law of Thermodynamics, the most probable consequence of an initial non-equilibrium state of a closed system is a steady increase of its entropy\cite{LL5}. This statement is frequently described by inequality $\dot{S}\ge 0$ which explicitly distinguishes past and future. Such asymmetry is usually referred to as the irreversibility of thermodynamics and contrasted with perfect time-reversal symmetry of mechanics (both classical or quantum). One of the objections (also known as Loschmidt's paradox) to Boltzmann's proof of $H$-theorem is based on this conflict: it should not be possible to single out a preferred direction in time from time-symmetric dynamics.

To resolve this difficulty we follow the arguments discussed by Bronstein, Landau\cite{bronstein}, and Peierls\cite{peierls_surprises_1979}. Consider equilibrium ensemble represented by multiple copies of the same closed system. Time-symmetric equations of motion determine the paths traversed by each realization. Projections of these paths to $(t,S)$ plane are the plots of time-dependent entropy $s(t)$. Fluctuations are responsible for deviations of $s(t)$ from equilibrium entropy $S_\text{max}$. The average entropy does not depend on time $\avrg{s(t)}=\const \lesssim S_\text{max}$.

\begin{figure}[ht]
\input{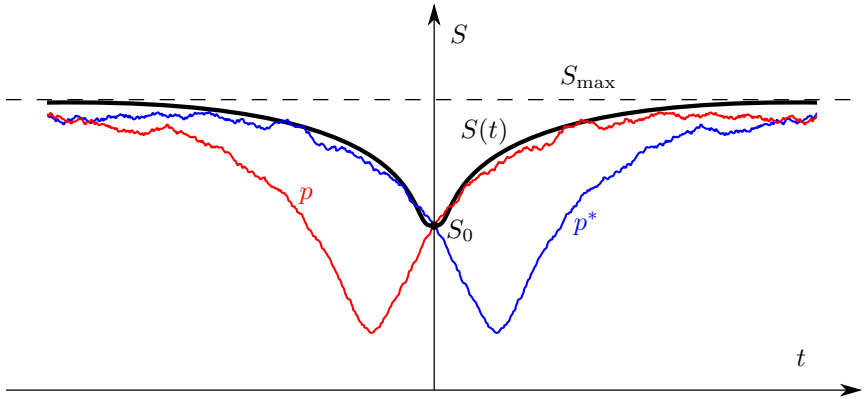}
\caption{A random ``forward'' path $\color{red}p$ with the entropy $\color{red}s(t)$, respective ``backward'' path $\color{blue}p^*$ with the entropy $\color{blue}s(-t)$, and the ``average'' entropy $S(t)$.}
\label{eplot}
\end{figure}

Now select a sub-ensemble corresponding at $t=0$ to the same entropy $s(0)=S_0<S_\text{max}$, see Fig.\ref{eplot}. These paths represent more or less improbable fluctuations, their projections to $(t,S)$ plane pass through $S_0$ at $t=0$. For each path $p$ (with the entropy $s(t)$) there exists a conjugate, time-reversed path $p^*$ (with the entropy $s(-t)$ --- the entropy itself is invariant under time reversal). The sub-ensemble is time-symmetrical and the probabilities of ``forward'' and ``backward'' paths are equal. The average plot is also perfectly symmetric. The conditional average (denoted below as $\avrg{\dots}_0\equiv \avrg{\dots}_{s(0)=S_0}$) of the entropy is not a constant any more $S(t)\equiv\avrg{s(t)}_{0}\ne\const$, but it is an even function $S(t)=S(-t)$ and has zero derivative at origin 
\begin{equation}\label{dot-s}
\dot{S}(0)=\avrg{\dot{s}(0)}_{0}=0.
\end{equation}
The Second Law in this picture would imply that the function $S(t)$ has a minimum at $t=0$. Local time-symmetrical formulation of the Second Law is therefore
\begin{equation}\label{ddot-s}
\ddot{S}(0)=\avrg{\ddot{s}(0)}_{0}\ge 0.
\end{equation}
Interestingly, such statement can be analyzed rigorously.

\section{Classical example}
Consider a simple thermalization problem: a closed system consists of two subsystems with initially different temperatures $T_1$ and $T_2$. Total Hamiltonian $H(\q,\p)=H_1(\q,\p)+H_2(\q,\p)$ gives conserved total energy and includes the interaction term. Generalized coordinates and momenta for entire system are denoted as $\q$ and $\p$. Initial partition corresponds to the local equilibrium
\begin{equation*}
\rho(\q,\p) = \frac{\exp(-N(\q,\p))}{\int \exp(-N(\q,\p))\D\q \D\p} \equiv
A \E ^{-N(\q,\p)},
\end{equation*}
where $A>0$ is the normalization constant and $N(\q,\p)$ is the ``weighted Hamiltonian''
\begin{equation*}
N(\q,\p)= \frac{H_1(\q,\p)}{T_1} + \frac{H_2(\q,\p)}{T_2}.
\end{equation*}
Similar setup is discussed in the paper by Jarzynski and W\'ojcik~\cite{jarzynski_classical_2004}, but in their scenario the interaction is turned on only temporarily. It is convenient to monitor the variation of the quantity $N(\q,\p)$, in a closed system it uniquely determines the energy (or heat) transfer between subsystems.

The dynamics of $N$ in the vicinity of the initial state (so that $T_1$ and $T_2$ are constant) should be identified with the entropy production, in agreement with \eqref{dot-s} its rate at origin is zero:\footnote{Transformations in \eqref{dot-s-class}  and \eqref{ddot-s-class} are based on Liouville's theorem: $\int \dot{f}(\q,\p) \D\q \D\p =0$ for arbitrary dynamic variable $f(\q,\p)$.}
\begin{multline}\label{dot-s-class}
\dot{S}=
\avrg{\dot{s}}_0 = \avrg{\dot{s}_1+\dot{s}_2}_0 = \avrg{\frac{\dot{H}_1}{T_1} + \frac{\dot{H}_2}{T_2}}_0 \equiv \avrg{\dot{N}}_0=
\int \dot{N} \rho \D\q \D\p =\\
A \int \dot{N} \E^{-N} \D\q \D\p=
-A \int \frac{\D}{\D t} \E^{-N} \D\q \D\p 
 =0.
\end{multline}

As expected from the Second Law \eqref{ddot-s}, the second derivative turns out to be non-negative
\begin{multline}\label{ddot-s-class}
\avrg{\ddot{N}}_0=
\int \ddot{N} \rho \D\q \D\p =
A \int \ddot{N} \E^{-N} \D\q \D\p=\\
-A \int \dot{N} \frac{\D}{\D t} \E^{-N} \D\q \D\p =
A \int \left(\dot{N}\right)^2 \E^{-N} \D\q \D\p
\ge 0.
\end{multline}

\section{Quantum-mechanical example}
Similar results can be obtained for quantum-mechanical evolution. Hamiltonians of two subsystems are both Hermitian, their sum is the total Hamiltonian $\hat{H}=\hat{H}_1+\hat{H}_2$.
The density matrix of the initial state again corresponds to the local equilibrium with temperatures $T_1$ and $T_2$
\begin{equation*}
\hat{\rho}= \frac{\exp(-\hat{N})}{\Tr \exp(-\hat{N})}\equiv
A \E^{-\hat{N}},
\end{equation*}
where $\hat{N}=\hat{H}_1/T_1 + \hat{H}_2/T_2$.
Note, that due to interaction $[H_1,H_2]\ne 0$ and the density matrix can not be factorized
\begin{equation*}
\E^{-\hat{N}} \ne \E^{-\hat{H}_1/T_1} \E^{-\hat{H}_2/T_2}.
\end{equation*}
The operator $\hat{N}$ is also Hermitian and has a complete set of eigenvectors, they are also eigenvectors for the density matrix:
\begin{equation*}
\hat{N}\ket{n}=n\ket{n},\qquad 
\hat{\rho}\ket{n}=A \E^{-n}\ket{n}.
\end{equation*}

We again start with a check that the time-reversal symmetry is not violated by the first derivative:
\begin{equation}\label{dot-s-quant}
\avrg{\hat{\dot{N}}}_0=
\Tr \hat{\rho} \hat{\dot{N}}=
\I \Tr \hat{\rho} [\hat{H},\hat{N}]=
\I \Tr \left(\hat{\rho}\hat{H}\hat{N} - \hat{\rho}\hat{N}\hat{H}\right)=
0.
\end{equation}

Averaging of the second derivative is slightly more complicated (here we designate $H_{nn'}=\bra{n}\hat{H}\ket{n'}$):
\begin{multline}
\avrg{\hat{\ddot{N}}}_0=
\Tr \hat{\rho} \hat{\ddot{N}}=
- \Tr \hat{\rho} \left[ \hat{H}, [\hat{H},\hat{N}]\right]=\\
- \Tr
	\left(
		 \hat{H}\hat{N}\hat{\rho}\hat{H}
		-\hat{\rho}\hat{H}\hat{N}\hat{H}
		-\hat{N}\hat{H}\hat{\rho}\hat{H}
		+\hat{\rho}\hat{N}\hat{H}\hat{H}
	\right)=\\
 -A\sum\limits_{nn'}
	\Big(
		 H_{nn'} n' \E^{-n'} H_{n'n}
		-\E^{-n}H_{nn'} n' H_{n'n}\\
		-n H_{nn'} \E^{-n'}H_{n'n}
		+\E^{-n} n H_{nn'} H_{n'n}
	\Big)=\\
 -A\sum\limits_{nn'}
	\left|H_{nn'}\right|^2
	\left(n' - n\right)
	\left(\E^{-n'} - \E^{-n}\right)\ge 0.
\end{multline}
This implies that the non-equilibrium state we consider is the local minimum of the entropy: it originates, on average, from a state with higher entropy and evolves, on average, into a state with higher entropy. In other words, at $t \gtrsim 0$ the energy, on average, flows from the hotter subsystem to the colder one.

We have divided the system into two parts only for convenience, the proof does not depend upon exact number of subsystems.
Obtained result can be applied to arbitrary spatial temperature distribution $T(\mathbf{x})$ by assigning
\begin{equation*}
\hat{N}=\int \frac{\hat{H}_\mathbf{x}}{T(\mathbf{x})}\D \mathbf{x},
\end{equation*}
where $\hat{H}_\mathbf{x}$ is the Hamiltonian density.

\section*{Acknowledgements}
I thank A.F.\,Andreev and O.A.\,Sudakov for fruitful discussions.


\begin{thebibliography}{9}
\bibitem{pit}{L.P.\,Pitaevskii
\textit{Phys. Usp.}
\textbf{54}, 625 (2011).
}
\bibitem{LL5}
{L.D.\,Landau, E.M.\,Lifshitz, \textit{Statistical Physics},
part 1 (Pergamon Press, Oxford, 1980).}

\bibitem{bronstein}
{
M.P.\,Bronstein, L.D.\,Landau, 
\textit{Phys. Z. Sowjet.}
\textbf{4}, 114 (1933); Collected Papers of Landau, edited by D.\ ter Haar, 69 (Pergamon Press, 1965).
}

\bibitem{peierls_surprises_1979}
{R.E.\,Peierls, \textit{Surprises in Theoretical Physics}, (Princeton University Press, 1979).}

\bibitem{jarzynski_classical_2004}
{C.\,Jarzynski, D.K.\,W\'ojcik, 
\textit{Phys. Rev. Lett.}
\textbf{92}, 230602 (2004).
}
\end{thebibliography}
\end{document}